# Understanding How Students Use Physical Ideas in Introductory Biology Courses

Jessica Watkins[1], Kristi Hall[2], Edward F. Redish[1], and Todd J. Cooke[3]

[1]Department of Physics, [2]Department of Curriculum and Instruction,
[3]Department of Cell Biology & Molecular Genetics
University of Maryland, College Park, MD 20742, USA

**Abstract.** The University of Maryland (UMD) Biology Education and Physics Education Research Groups are investigating students' views on the role of physics in introductory biology courses. This paper presents data from an introductory course that addresses the fundamental principles of organismal biology and that incorporates several topics directly related to physics, including thermodynamics, diffusion, and fluid flow. We examine how the instructors use mathematics and physics in this introductory biology course and look at two students' responses to this use. Our preliminary observations are intended to start a discussion about the epistemological issues resulting from the integration of the science disciplines and to motivate the need for further research.

**Keywords:** curriculum reform, physics in biology, epistemology
**PACS:** 01.40.Ha, 01.40.Di, 01.40.Fk

## INTRODUCTION

For over a decade, researchers, policy-makers, and educators have advocated for the reform of life science education [1-4]. In 2003, the National Research Council (NRC) issued *Bio 2010* [1] advocating the need to transform undergraduate biology education. More recently, similar reports were published by the Howard Hughes Medical Institute, American Association of Medical Colleges, and the Board on Life Sciences [2][3]. All of these broad-based initiatives emphasize two goals for reform of biology instruction: (1) greater incorporation of chemistry, physics, and mathematics and (2) the development of sophisticated scientific competencies and skills, including quantitative reasoning and the application of fundamental physical and chemical principles to biological processes.

In light of these calls for reform, more research needs to be conducted to understand the implications of the integration of the different science disciplines. In particular, we need to better understand (1) the epistemological differences between the disciplines, (2) students' own ideas about these distinctions, and (3) given these, the impact of integration of the sciences. In this paper, we argue for the need for this in-depth research, involving both physics education and biology education researchers. To motivate this work, we discuss some of the initial epistemological issues we observe emerging from the reform of an introductory biology course. We discuss our preliminary observations on how biology instructors present equations to the students, comparing their approach to what is typically found in a physics course. We also present preliminary interview evidence to draw attention to some of the different epistemological issues that biology students may face. *Our results are by no means conclusive, but are intended to start a conversation about the need for research at the boundaries between the science disciplines.*

## BACKGROUND

Understanding discipline-specific epistemologies has far-reaching implications for both instructors and students. Previous research suggests that students bring in previous ideas and expectations about the nature of the knowledge they are learning [5]. These ideas can include misunderstandings about what counts as knowing and understanding, about what kinds of knowledge and learning their courses are trying to teach, and about what is appropriate for them to do to learn it. And just as students' preconceptions about content can hinder their learning, students' epistemological ideas can constrain their approach to learning, even in reformed classrooms.

In physics education, researchers have documented that student views about physics knowledge, e.g., as formulas rather than as concepts expressible in equations, negatively affect their approaches to learning [6]. Similarly, students who view biology as a set of disconnected facts to be absorbed and regurgitated may view the principles and equations used by their professor as additional information to be memorized, not as tools to construct a deeper understanding of biology. While we can hypothesize about biology students' epistemologies based on anecdotal evidence, little research has been done to document student ideas about biology knowledge and learning, much less on how these ideas interact with the reforms underway in their introductory courses.

For this paper, we are interested in exploring issues arising from the integration of physics and mathematics in introductory biology courses. While a great deal of research has been conducted in understanding biology students' ideas about physics and math in the context of a physics course [7], the issues that arise in a biology course may be of a very different nature. To understand the epistemological issues that impact biology students, particularly those that relate to the incorporation of physics and math in introductory biology courses, we need both a better picture of how both biology instructors and biology students view the nature of biological knowledge and learning.

## DESCRIPTION OF COURSE

The introductory biology course under study, Organismal Biology (Org Bio), concentrates on the diversity, structure, and function of all organisms. The traditional approach to teaching this course is almost universally derided by both instructors and students as a "forced march through the phyla." The fundamental principles governing the diversity, structure, and function of all organisms often do not emerge from the tsunami of isolated organism-specific facts.

The biology faculty at UMD reformed Org Bio to focus on basic principles in biology, chemistry, and physics, and their implications for organisms. These resulted in the identification of several principles that serve as the organizational framework, including the relevance of universal physical and chemical laws and importance of diverse structure-function relationships.

Physical and chemical principles play a critical role in the new curricula, as students are asked to consider how life is governed by these laws. In particular, Org Bio emphasizes understanding how living organisms have exploited universal physicochemical principles to evolve diverse structure-function relationships for carrying out life's fundamental processes, such as gas exchange, motility, and nutrient assimilation.

In addition to curricular reform, the instructors are also making great strides to reform the pedagogy of Org Bio. The instructors incorporated several active-engagement activities, with the goals of teaching the students not just content, but also how to approach the principle-based concepts. To accomplish these goals, the instructors made their lectures more interactive, with clicker questions and class discussions. They also dedicated a third of the class periods to small group activities, incorporating concept-mapping, group and class discussions, and enactments.

## METHODS

To gain a sense of how biology students use physical ideas in this course, we focused on both the instructors' presentations and the students' responses. For this paper we will rely on our qualitative data sources, including field notes from classroom observations and student interviews.

To collect field notes, two researchers observed more than half of the lectures during the semester, writing descriptive narratives and rough transcriptions of the instructors' presentations. The field notes focused specifically on how the instructors presented the use of physics and equations in the context of the biology. Eleven student interviews were conducted throughout the semester, centering on the different conceptual and epistemological issues of the course as found in the field notes.

## USE OF EQUATIONS IN ORG BIO

By examining how two biology instructors reform their course, we begin to gain insight into how biologists integrate physics and mathematics in their introductory courses. Here we detail how they treat equations and facilitate quantitative reasoning for introductory biology students. We describe our field notes from a lecture in the Org Bio course that directly addressed equations—both the concepts behind the equations as well as how to use the equations in the course.

The instructor first wrote the diffusion equation on the board:

$$J = -D \frac{\Delta C}{\Delta x}, \qquad (1)$$

and asked students to remember and call out that $J$ is the diffusion rate, $D$ is the diffusion coefficient, $\Delta C$ is the change in concentration, and $\Delta x$ is the distance. *"Whenever you see an equation like this, think about what happens when you change a given variable."* The instructor then asked what happens to the rate of diffu-sion when you make a given change to the biological system. When the students were silent for a while, the instructor modified the question to indicate what the students should be thinking about: *"Leave the equation to the side. What's your intuition [tell you]?"* The instructor and students then spent a significant portion of the rest of the lecture exploring factors that affect the rate of diffusion and maximum thickness through which a gas can diffuse, using the equations as referents and checking with their intuition and knowledge of biological systems. Later in the discussion the instructor reiterated: *"Again, when you see an equation, what happens when you change the variables?"*

The above scenario could easily be seen in a physics classroom, simply replacing the biology words with those for a physical system. Students are asked to draw on similar conceptual and epistemological ideas as the instructor promotes mapping the equation to the physical and biological systems they represent. However, how this biology course *applies* quantitative reasoning appears to be different than what is typically found in an introductory physics course. In this reformed biology course, the instructor often returned to the equations during a discussion of form and function of organisms or when talking about the affordances and constraints of physical laws in the evolutionary story. For example, the diffusion equation was brought up again when students were asked to devise a model organism with certain specifications, e.g., small size, fast-moving, underwater. Given these constraints, students had to consider the physical laws previously discussed to plan how this model organism would function. In contrast, the use of equations in introductory physics courses most often leads to a problem-solving activity with a precise answer. The equation is not only used as a tool to provide insight on a physical situation or describe the 'rules' of physics, but also to perform calculations or connect different physical ideas through mathematics. We speculate that differences between the disciplines may not occur in the actual process of quantitative reasoning, but when asking "to what end?" From these data, we cannot generalize beyond this one course, but understanding how biologists use equations to tell the story of how an organism functions or evolves will better inform how to integrate the disciplines in a deeper way.

## STUDENT RESPONSE TO EQUATIONS

After this lecture, we interviewed students to examine their response to the use of physical equations in their biology course. We asked students about their thoughts on the use of equations in the course and the role of equations in biology. Here we present two different student responses.

## Biology Is More than Letters and Numbers

One of the students we interviewed, Ashlyn, had a very strong reaction to the use of equations in the course. After the interviewer asked about the recent use of equations in the group activities and lecture, Ashlyn gave a very negative response:

*"I don't like to think of biology in terms of numbers and variables. I feel like that's what physics and calculus is for. So, I mean, come time for the exam, obviously I'm gonna look at those equations and figure them out and memorize them, but I just really don't like them."*

Her first comments about equations were centered on her negative affective response, i.e. how she did not like the equations in the course. She also talked about the roles of the different disciplines in using "numbers and variables" to describe phenomena, suggesting that these representations do not belong in biology. When probed more about how she studies equations in this course, she elaborated:

*"It's memorizing how they fit together. If you give me, like, for example, like, the diffusion equation on the last exam, if you gave me the units, I could figure it out for the most part, but the equations with the letters that stand for numbers, sometimes I can't remember which letters stand for what.. it's basically a way to put it, put the concept into words. I think that's what the only function of the equations are."*

Ashlyn again lamented the use of letters and numbers, but explained how she can use the units or concepts to explain the relationships described by the equations. She reported that she memorized the equations in terms of letters and numbers, but found that to be less useful than thinking about the concepts. Interestingly, she did not appear to be viewing equations as devoid of physical meaning, in contrast to findings in introductory physics students [5]. Instead, she seemed to be responding negatively to the use of equations as referents for biological phenomena, which is how the instructors primarily used equations in the course. Her later comments shed light on this issue:

*"I think that biology is just—it's supposed to be tangible, perceivable, and to put that in terms of letters and variables is just very unappealing to me, because like I said, I think of it as it would happen in real life, like if you had a thick membrane and you try to put some-*

*thing through it, the thicker it is, obviously the slower it's gonna go through. But if you want me to think of it as 'this is x' and 'that's d' and then 'this is t,' I can't do it. Like, it's just very unappealing to me."*

The way in which the instructors used the equations to refer to the underlying physical laws conflict with Ashlyn's thoughts about what biology is "supposed to be." Her ideas about the nature of biology, specifically that it is "tangible" and "perceivable," suggest that there are unique challenges for the incorporation of physics and mathematics in biology courses. Her interview again points to the need for better understanding of the epistemologies and student expectations of the different science disciplines.

## Equations as tools for sense-making and communication

May was also a student in Org Bio, who offered a different response to the use of equations in the course. She reported a much more positive affective response—*"I like equations. It helps me."*—and freely discussed the ways in which equations were useful:

*"It's just scientists trying to understand one... like make sense of one specific aspect [of life], which is diffusion in this case... and we have to have universal codes for things, so that we can talk about them and make sense of them."*

Interestingly, May did not make a distinction between "scientists" here, unlike the differences between biology and physics that Ashlyn brought up. She talked about why scientists would use equations, touching on the role of equations in sense-making and communication. After she elaborated on the use of equations in communication, the interviewer asked about the differences between an equation and a paragraph to communicate scientific ideas. May responded:

*"I don't know, maybe this is more memorization. You memorize the equation and then... With people who learn with equations, they have the equation and then they understand it. As opposed to understanding it with the paragraph and trying to figure out things from there. Like you could... start with an equation and work with the different variables in it... and then from there you understand, oh, the rate goes up because surface area went up…"*

Similar to Ashlyn, May reported viewing equations as representing physical or biological concepts, such as diffusion rate and surface area. May appeared to see equations as tools for sense-making—something to memorize and then use to understand relationships. To May, the usefulness of the equation was in its simplicity and, given the vast amount of information that biology students are required to assimilate, this feature of equations may be especially appealing.

## IMPLICATIONS FOR PER

While we are just beginning this research, our observations of a reformed biology course and interviews of students suggest that efforts to integrate physics and mathematics into biology courses will require collaboration between biologists, physicists and other scientists. Based on the preliminary issues that arose in our investigations, we believe there is a need for greater understanding of the epistemological differences between the disciplines themselves and the epistemological ideas that students hold for each. Understanding these differences requires significant contributions from the physics education community, for disciplinary perspective as well as the tools and analytic methods to inform the transformation of undergraduate biology education. Furthermore, efforts to reform physics courses to incorporate more biology may also encounter similar epistemological issues.

## ACKNOWLEDGMENTS

The authors wish to acknowledge the instructors of the Org Bio course and the Biology and Physics Education Research Groups at the University of Maryland for their important contributions. This work was supported by NSF grant 09-19816.

## REFERENCES

1. NRC, *Bio 2010: Transforming Undergraduate Education for Future Research Biologists*, NAP, 2003
2. AAMC-HHMI committee, *Scientific Foundations for Future Physicians*, AAMC, 2009
3. NRC, *A New Biology for the 21st Century: Ensuring the US Leads the Coming Biology Revolution*, NAP, 2009.
4. J. B. Labov, A.H. Reid, and K.R. Yamamoto, "Integrated Biology and Undergraduate Science Education: A New Biology Education for the 21st Century?" *Cell Biology Education,* **9**, 10-16 (2010).
5. E.F. Redish, J.M. Saul, and R.N. Steinberg, "Student Expectations In Introductory Physics," *Am. J. Phys.,* **66**, 212-224 (1998).
6. L. Lising and A. Elby, "The impact of epistemology on learning: A case study from introductory physics," *Am. J. Phys.*, **73**, 372-382 (2005).
7. T.L. McCaskey, *Comparing and contrasting different methods for probing student epistemology and epistemological development in introductory physics* (unpublished dissertation) Physics, UMD, (2009).